# THE INFLUENCE OF COLLABORATION IN PROCUREMENT RELATIONSHIPS

Wesley S. Boyce[1], Haim Mano[2] and John L. Kent[3]

[1]Department of Management Sciences, University of Iowa, Iowa City, Iowa
[2]Marketing Department, University of Missouri – St. Louis, St. Louis, Missouri
[3]Department of Supply Chain Management, University of Arkansas, Fayetteville, Arkansas

## ABSTRACT

*Supply Chain Management often requires independent organizations to work together to achieve shared objectives. This collaboration is necessary when coordinated actions benefit the group more than the uncoordinated efforts of individual firms. Despite the commonly reported benefits that can be gained in close relationships, recent research has indicated that collaboration attempts between purchasing firms and their suppliers have not been as widespread as anticipated. Using a survey of procurement professionals, this research investigates how the purchasing function utilizes collaboration in its supply chain relationships. Structural equation modeling is used to identify how information sharing, decision synchronization, incentive alignment, collaborative communication, and trust impact collaboration, as well as how collaboration impacts performance. Results from 86 survey responses indicate that firms are still not fully utilizing collaborative relationships.*

## KEYWORDS

*Collaboration, supply chain relationships, procurement, purchasing, structural equation modeling*

## 1. INTRODUCTION

Supply chain management (SCM) has evolved to a point where collaboration is common practice for firms to achieve shared objectives. This has evolved from a time when firms expressed concern only about their own interests and other firms were leveraged rather than cooperated with. As noted by Mentzer et al (2001), many definitions exist that try to capture the essence of SCM, and a majority of these definitions incorporate or even require supply chain collaboration (SCC) as a key component. As a result, many firms are now making an effort to collaborate with supply chain partners.

Purchasing firms have utilized numerous tools in their efforts to work more closely with suppliers. Key to these efforts are dimensions of collaboration that facilitate close relationships. These include information sharing (Simaputang and Sridharan, 2004), decision synchronization (Stank et al, 2001; Simaputang and Sridharan, 2004), incentive alignment (Manthou et al, 2004; Simaputang and Sridharan, 2004), goal congruence, resource sharing, joint knowledge creation, and collaborative communication (Cao and Zhang, 2011). These dimensions are often interrelated, and causal relationships may exist between them (Cao and Zhang, 2011).

Collaboration is enhanced by utilizing these dimensions since they bring the interest of the supply chain to the forefront rather than that of any individual firm.

Numerous benefits have been outlined in the literature that rationalize the choice to engage in collaborative relationships. Firms participating in collaboration have an opportunity to be more efficient (Kalwani and Narayandas, 1995), more customer focused by exchanging information about customer needs (Myers and Cheung, 2008), and more successful overall than those not participating (Kalwani and Narayandas, 1995; Simatupang and Sridharan, 2004). Sales growth, market share, and satisfaction often increase, and working closely together makes firms more likely to extend their partnerships into the future (Ramanathan and Gunasekaran, 2014). Supply chains may even become more resilient by managing risk as a network rather than at the firm level (Christopher and Peck, 2004). Despite these benefits, many firms have struggled to engage in collaboration due to struggles with partner selection and matching the needs and goals of independent organizations (Daugherty et al, 2006). Firms have also struggled to identify who to collaborate with, and a lack of trust between partners has been an issue (Barratt, 2004). Additionally, the decision to engage in a collaborative relationship requires commitment from all involved parties since collaboration efforts can lose momentum when faced with resistance (Fawcett et al, 2015).

Although close supply chain relationships may have major potential benefits, it is important to note that not all relationships should be collaborative in nature and collaboration is not appropriate in all situations (Lambert et al, 1996). For example, previous literature has indicated that integrating with suppliers may lead to poorer quality outputs or stifle innovation for some firms (Koufteros et al, 2005; Swink et al, 2007). Thus, collaboration should only be considered as a strategy to approach when the benefits of working together outweigh the costs (Terjesen et al, 2012). In other words, there will be times where an arms-length relationship is appropriate, such as with items that are not strategically important to a firm, and others where an intimate link is appropriate. This is consistent with the findings of Golicic et al (2003), who note that firm relationships should have varying levels of magnitude.

The research question to be considered in this paper is whether or not collaboration influences relationships between procurement and suppliers, or as Mentzer et al (2001) stated it, how prevalent is supply chain management from the perspective of procurement? We seek to discover whether buyers at companies in a supply chain become involved in immersive relationships with channel partners that are full of trust and knowledge sharing, or are relationships still combative whereby each firm is solely interested in its own well-being? While firms have likely found a middle ground between their own success and that of the supply chain, outlining the current state of collaboration can provide insights into how relationships have developed since the time when contentious relations were common and how they need to continue to evolve. We contribute to the current knowledge of collaboration by presenting a view of how procurement utilizes the strategy. We also provide an updated perspective of how firms view their collaboration efforts, which is crucial for a strategy that has been considered to be critical to SCM.

The remainder of the paper is structured as follows. First, the theory that SCC is grounded in is discussed, followed by the conceptual framework and a review of key pieces of literature that involve collaboration-related studies. Next, the hypotheses and methods used for empirical testing of the proposed framework are explained, followed by results of the study. The paper ends with conclusions, limitations, and suggestions for additional research.

## 2. THEORETICAL FOUNDATIONS

The literature outlines three major perspectives to classify SCC, including transactional, relational, and resource-based. Powell (1998) notes that research on SCC has focused primarily on either a transactional, exchange oriented focus or a more relational, process-based focus. While the transactional view serves as a coordinating mechanism both within and outside the firm (Williamson, 1975), the relational view focuses on multi-firm efforts that seek to benefit the supply chain as a whole (Powell, 1998). In addition, authors like Jap (1999) note a resource-based view as also being critical, and Lavie (2006) expands upon that by mentioning an extended resource-based view whereby collaborating firms combine resources to gain a competitive advantage (Cao and Zhang, 2011).

### 2.1 Transactional View

In his work on transaction cost economics (TCE), Williamson (1975) proposes two methods of organization that involve market transactions and hierarchy transactions. Market transactions are those that support the coordination of buyers and sellers and involve firms conducting business with those companies that offer the most attractive terms, such as price. Hierarchy transactions support coordination within the firm and include initiatives like vertical integration. Koh and Venkatraman (1991) suggest a third method of organization for SCC that helps to avoid these factors. Their study of joint ventures and how they impact the market value of parent firms showed that this SCC organization method can limit costs related to opportunistic behaviors and monitoring partners in market transactions (Croom, 2001). This approach can also negate the limiting factor of hierarchy transactions since they may not be effective when a firm is forced to internalize an activity that does not match its competencies (Cao and Zhang, 2011).

### 2.2 Relational View

Competitive advantage can result as firms focus more on working together. This is relevant to SCC since the strategy requires firms to work closely to achieve mutual goals. The relational view builds upon the resource-based view by expanding the previously mentioned critical resources beyond firm boundaries to create joint profits from working in tandem than those that could be generated individually (Dyer and Singh, 1998). The key to this view is that the firms involved are able to generate benefits together that they would be unable to generate in isolation (Cao and Zhang, 2011) and long-term profits are based on network relations (Duschek, 2004). Thus, firms have an incentive to work together for mutual benefit in the form of long-term profitability.

### 2.3 Resource-Based View

The resource-based view (RBV) begins with a firm gauging its key assets, including its own resources, capabilities, and core competencies (Barney, 1991; Japp, 1999). Porter (1985) notes this resource-based view and how it can lead to a competitive advantage when a firm utilizes its resources and capabilities more effectively than its rivals. While Porter's and other early research on this topic considered the tangible and intangible assets of the firm, Dyer and Singh (1998) note that these resources may extend beyond firm boundaries and be a part of interorganizational processes. More specifically, they claim that firms that combine resources in unique and difficult to imitate ways may realize a competitive advantage over other firms that are unable to do the same. This perspective helps to facilitate collaboration by giving firms the opportunities to focus on what they do best and allowing partners to handle the rest, which can also improve the competitive position of a firm or group of firms.

## 3. CONCEPTUAL FRAMEWORK AND HYPOTHESIS DEVELOPMENT

The theoretical foundations of collaboration are enabled by interfirm activities that promote SCC, and the previously noted dimensions aid purchasing firms and supply chains in gaining a collaborative advantage. This research investigates the impact of information sharing, decision synchronization, and incentive alignment on buyer/supplier relationships. While this is not an all-inclusive list of the dimensions that are relevant to SCM, these are commonly encountered in the literature. In addition, while there are other issues that are critical aspects of SCC, most notably referring to trust and commitment, these are encompassed within and a significant aspect of these dimensions.

### 3.1 Information Sharing and Collaborative Communication

Information sharing is a frequently noted dimension of SCC since shared information forms the backbone of interfirm relationships. It enables each of the theoretical constructs mentioned above since it facilitates the exchange of data regarding sales, customer needs, market structures, and demand levels (Myers and Cheung, 2008). Potential benefits of shared information include a reduced incidence of the bullwhip effect, early problem detection, faster response, and trust building (Lee and Whang, 2001). For example, Kwon and Suh (2004) note that information sharing reduces uncertainty levels and thereby improves the degree to which firms trust one another. This is a key aspect of SCC because shared information facilitates firms' ability to meet end user needs (Spekman et al, 1998) and free exchanges of information have been found to be very effective in reducing the risks of supplier failure (Lee, 2004). Collaborative communication can enhance the sharing of information since it facilitates the transmission of information between firms (Cao and Zhang, 2011).

### 3.2 Decision Synchronization

Decision synchronization is a dimension of SCC that has the potential to reduce a source of conflict inherent in supply chain relationships. As defined by Simaputang and Sridharan (2002), it is the degree to which channel partners are able to coordinate critical decisions in planning and operations that benefit the supply chain as a whole. It has been noted that this dimension impacts information sharing, but it also has an effect on incentive alignment since different channel members are responsible for different types of decisions (Simaputang and Sridharan, 2005). Therefore, decision synchronization helps to facilitate incentive alignment, which allows firms to appropriately devise incentives based on the level of responsibility a party owns.

### 3.3 Incentive Alignment

Incentive alignment is another vital dimension of SCC since an underlying necessity of collaboration is to have common goals. Aligned incentives reduce the incidence of channel partners making decisions that are limited to their own benefit. Simaputang and Sridharan (2002) describe it as a way to share costs, benefits, and risks across all supply chain partners. They note multiple benefits from this dimension, including improved commitment between partners and a foundation from which to build trust since firms are considering the benefit of the supply chain rather than just their own. Finally, it has been noted that gains and risk should be shared equitably to ensure they are fair in regards to the level of level of investment and risk a firm is accountable for (Lee and Whang, 2001; Manthou et al, 2004).

## 3.4 Comparing Dimensions of Collaboration

With these dimensions of collaboration in mind, the degree to which firms are working together is also of interest. Do firms simply exchange the required information in order to conduct business or are they sharing resources and developing close, long-term relationships to improve the supply chain? The previously discussed dimensions of collaboration provide insights into these different levels of depth in collaborative relationships. Therefore, hypotheses 1a through 1e address the issue of the intensity of supply chain relationships in the current environment. Namely, it is expected that firms are more likely to implement the lowest levels of the continuum before implementing it at higher levels. For example, information can be shared in different degrees, so it is expected that information sharing should be more prevalent than the other dimensions.

**H1a**: Firms are more likely to implement the practice of information sharing than collaborative communication.
**H1b**: Firms are more likely to implement the practice of information sharing than decision synchronization.
**H1c**: Firms are more likely to implement the practice of information sharing than incentive alignment.
**H1d**: Firms are more likely to implement the practice of collaborative communication than decision synchronization.
**H1e**: Firms are more likely to implement the practice of collaborative communication than incentive alignment.

## 3.5 A Structural Model

A conceptual model of the relationship between collaboration practices and improved firm performance is depicted in Figure 1. Since the literature review presents evidence that fully immersive collaborative partnerships are rare, the model posits that implementing a degree of collaboration in supply chain relationships should lead to improved performance. Thus, while firms in a supply chain may not practice business as a single entity in the true spirit of collaboration, they may still gain an advantage from participating in collaborative practices like sharing forecasts or including partners in the product design process. The model also takes into account the dimensions of collaboration that firms can adopt to further their advancement towards the practice of SCM, which also enhance collaboration practices and ultimately improve performance.

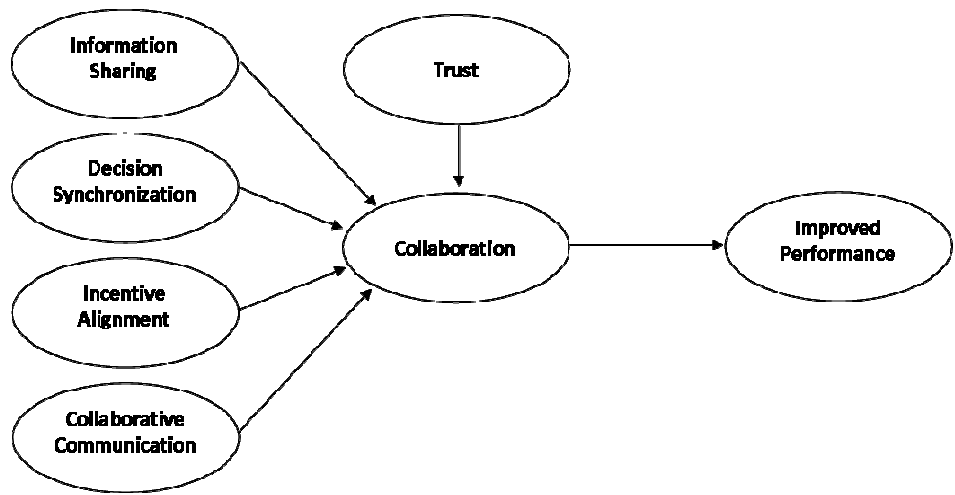

Figure 1. Conceptual model

The literature places much importance on these dimensions and their contribution to collaboration. For example, information sharing has been identified as one of the core parts of a collaborative business model (Fawcett et al, 2007). In addition, the sense of belonging and aim of a common goal associated with decision synchronization, as well as the knowledge that aligning incentives for the good of the supply chain can help to ensure future survival should both lead to a positive focus on collaboration (Simatupang and Sridharan, 2004). Thus, it would seem that each should have a positive relationship with collaboration and lead to better relationships between firms. Hypotheses H2a through H2d address this by positing these relationships.

**H2a**: Information sharing is positively related to collaboration.
**H2b**: Decision synchronization is positively related to collaboration.
**H2c**: Incentive alignment is positively related to collaboration.
**H2d**: Collaborative communication is positively related to collaboration.

As previously mentioned, trust is a required key element in a long-term collaborative relationship (Sahay, 2003). When firms are working closely together and sharing potentially sensitive information, they need to have the confidence that their partner will not behave opportunistically. Firms must also understand that they have the responsibility to be mindful that the knowledge they gain from partners is private and not to be shared. Lewis (2000) found that trust is the primary concern firms have related to why supply chain relationships are not working well. Thus, trust is necessary not only for a collaborative relationship to exist, but for it to thrive. Hypothesis 3 posits this important role of trust on collaboration.

**H3**: Firms that report higher levels of trust with their channel partners will also report higher levels of performance from their collaborative relationships.

Collaboration can lead to benefits like greater visibility, reduced variability, and increased velocity in the supply chain. This greatly reduces the likelihood that problems like the bullwhip effect will arise (Lee and Whang, 2001) and leads to a level of competence that can make one supply chain dominant over its competitors. More specifically, Bowersox et al (2000) note how decision synchronization increases performance in the areas of on-time delivery and maintaining product availability. It has also been noted that information sharing leads to better performance in a supply chain (Lee et al, 1997; Whipple et al, 2002) and incentive alignment motivates channel partners to make decisions to ensure profitability of the supply chain is optimized (Simatupang and Sridharan, 2002). Therefore, firms should report that participation in collaborative relationships or practicing certain collaborative initiatives have led to numerous benefits and overall firm and supply chain performance should be improved by utilizing these strategies.

**H4**: Firms that report higher levels of collaboration with their channel partners will also report higher levels of performance from their collaborative relationships.

## 4. METHODOLOGY

This study utilizes an internet survey to gain the input of purchasing professionals to determine where the function as a whole stands in its collaborative efforts. Spekman et al (1998) demonstrate that purchasing managers have a critical role to play as their organizations transition through the SCM continuum.

### 4.1 Survey Instrument

The survey instrument was developed by conducting a review of the literature and scales were developed by utilizing Churchill's (1979) framework for construct development. This involves utilizing existing measures whenever possible and providing rationale for the development of new constructs. The survey instrument is comprised of questions that gauge respondents' opinions of their firm's utilization of collaboration and is centered on the four previously mentioned dimensions of collaboration (see Appendix).  A 5-point Likert scale was used to indicate the level of agreement purchasing professionals had with each statement.

The survey instrument involved scales that measured the constructs outlined in the hypotheses. There were four scales that aimed to measure the degree to which respondent firms were involved in the previously mentioned dimensions of collaboration. Scales representing how performance and trust were impacted by collaboration were also included. Background variables, such as firm size or annual sales, were developed to see if differences existed between different types of firms. The questionnaire utilized in the research was pretested for content validity by having supply chain academics and professionals review it in order to ensure questions are precise, accurately worded, and understandable by the target audience.

### 4.2 Subjects

The survey was sent to purchasing professionals that were members of a large professional purchasing organization via its monthly newsletter. 97 surveys were submitted and 11 of them were omitted due to being incomplete. Therefore, the final sample size was 86 completed surveys.

Respondent backgrounds varied across the sample. The majority classified themselves as a buyer or manager at their respective firm. Most had a formal supplier agreement in place, and this was most commonly in the form of a contract. Nearly half of the firms had over 500 employees, and most had an annual sales volume of less than $100 million or greater than $250 million. Annual purchasing volume followed a similar pattern. Finally, nearly all participants reported working for private firms, with only a few working at public or government-related organizations.

Non-response bias was tested by comparing the first half of respondents to the second half by using t-tests of a random group of constructs. The results indicate no significant difference between the groups. In addition, firms in each group were similar in terms of sales volume, contract length, and number of employees. Thus, it does not seem that non-response bias was an issue in this research.

## 5. RELIABILITY AND FACTOR ANALYSIS

Reliability and factor analyses were conducted to ensure the scales were measuring what they were intended to measure and that each scale did in fact represent a single factor. Results can be seen in Table 1. Principal components factor analysis was conducted using principal components analysis procedure.

Table 1. Reliability and Factor Analytic Data

| Scale | Cronbach's Alpha | Kaiser-Meyer-Olkin Adequacy |
|---|---|---|
| Collaboration Continuum | 0.867 | 0.828 |
| Information Sharing | 0.882 | 0.749 |
| Decision Synchronization | 0.846 | 0.802 |
| Incentive Alignment | 0.840 | 0.857 |
| Collaborative Communication | 0.860 | 0.834 |
| Performance Improvement | 0.889 | 0.843 |
| Trust | 0.824 | 0.722 |

Initially, all but two scales loaded onto a single factor using the principal components procedure. This included the information sharing and trust scales. The information sharing scale had an initial alpha value of 0.857 and loaded onto two factors. However, eliminating one question allowed the alpha value to increase to 0.882 and led to a single factor. Similarly, the trust scale had an initial alpha value of 0.795 and loaded onto two factors, but it was clear that eliminating one question would allow the alpha value to increase to 0.824 and lead to a single factor for the scale. An item was also removed from the collaboration continuum scale since it led to an improvement in the alpha value from 0.861 to 0.867.

Cronbach's alpha was used to measure internal consistency of each of the scales. The reliability of all scales was acceptable according to widely accepted guidelines, which indicate that the alpha value should be at least 0.7 (Flynn et al, 1990). In addition, construct validity was confirmed using Flynn et al's (1995) example since each scale had items that all loaded on a common factor and the eigenvalues are all well above the threshold of 1.

Bivariate correlations were analyzed to see the relationships that exist between each of the survey scales, and this analysis indicates significant relationships exist between all of the scales. These can be seen in Table 2. The correlations of the study's scales with background variables were also investigated. The results indicate minimal interference from background factors, and the scales did not measure unintended constructs.

Table 2. Scale intercorrelations

| Scale | Collaboration Continuum | Information Sharing | Decision Synchronization | Incentive Alignment | Collaborative Communication | Performance Improvement | Trust |
|---|---|---|---|---|---|---|---|
| Collaboration Continuum | - | 0.653** | 0.579** | 0.511** | 0.537** | 0.372** | 0.625** |
| Information Sharing | | - | 0.567** | 0.505** | 0.560** | 0.336* | 0.549** |
| Decision Synchronization | | | - | 0.741** | 0.732** | 0.352** | 0.509** |
| Incentive Alignment | | | | - | 0.696** | 0.464** | 0.423** |
| Collaborative Communication | | | | | - | 0.370** | 0.575** |
| Performance Improvement | | | | | | - | 0.242* |
| Trust | | | | | | | - |

** Significant at $p < 0.01$

* Significant at $p < 0.05$

# 6. HYPOTHESES TESTING

## 6.1 Hypothesis 1

The group of H1 hypotheses compares the 4 dimensions of collaboration and proposes that there will be a statistically significant difference between them. This seeks to provide evidence that some dimensions are more commonly practiced than others. These hypotheses were tested using the scales that measured the dimensions of collaboration, including information sharing, decision synchronization, incentive alignment, and collaborative communication. The analysis of variance with repeated measures design was used to identify if there was a difference in collaboration practice based on the dimensions.

Identically worded questions were compared to determine whether respondents thought that their firms frequently engage in the dimension under examination and whether or not it was important. Results for the first comparison are listed in Table 3. This indicates that the mean value for how often information is shared is much higher than the other dimensions, with decision synchronization and incentive alignment having especially low values. The resulting F-value determined that the mean collaboration practice differed between dimensions (F = 68.333, p < 0.01).

Table 3. Analysis of variance with repeated measures: collaboration engagement

| Descriptive Data | Mean | Standard Deviation | Mauchly's Test | Chi-Square | Significance |
|---|---|---|---|---|---|
| Information Sharing | 3.52 | 1.024 | | 5.265 | 0.384 |
| Decision Synchronization | 2.19 | 0.988 | | | |
| Incentive Alignment | 2.24 | 1.060 | **Within-Subjects Effects** | **F** | **Significance** |
| Collaborative Communication | 3.18 | 1.055 | | 68.333 | 0.000 |

The second set of questions indicated the importance of each dimension from the perspective of respondents, and the analysis of variance with repeated measures information for that comparison is located in Table 4. The mean values for information sharing and collaborative communication are again higher than those of decision synchronization and incentive alignment. Mauchly's test was significant (chi-square = 13.125, p = 0.022), so evidence exists that the sphericity assumption was violated. Therefore, the Huynh-Feldt correction was applied since the Epsilon value was well above 0.75. This led to the conclusion that there were significant differences between the dimensions in their use of collaboration (F = 15.288, p < 0.01).

Table 4. Analysis of variance with repeated measures: collaboration importance

| Descriptive Data | Mean | Standard Deviation | Mauchly's Test | Chi-Square | Significance |
|---|---|---|---|---|---|
| Information Sharing | 3.35 | 1.192 | | 13.125 | 0.022 |
| Decision Synchronization | 2.78 | 1.248 | | | |
| Incentive Alignment | 2.81 | 1.200 | **Within-Subjects Effects** | **F** | **Significance** |
| Collaborative Communication | 3.55 | 1.160 | | 15.288 | 0.000 |

The results for the second repeated measures procedure indicate that respondents found information sharing and collaborative communication to be more important than either decision synchronization or incentive alignment. Therefore, the group of H1 hypotheses is not rejected since the evidence outlined above support those claims.

## 6.2 Hypotheses 2, 3, and 4

The final three hypotheses stated that the dimensions of collaboration and trust are positively related to collaboration practice, and that higher levels of collaboration will lead to improved performance. Structural equation modeling (SEM) was conducted in IBM SPSS Amos Version 21 software using the process outlined by Brunch (2008). The model consisted of two parts, including a structural model and a measurement model (Tan, 2001). The structural model outlines the causal relationships between the latent variables, while the measurement model deals with the constructs and their ability to measure the latent variables. Each of the scales represented one of these latent variables, which are unable to be measured directly but rather by a series of questions that comprised the scale. These questions served as the manifest variables for the model.

Initial fit results of the structural model indicated that there was indeed a good fit. The chi-square value of 5.455 (p = 0.363) is insignificant, which is one piece of evidence supporting a good fit. In addition, the CFI value of 0.998 is well above the suggested limit of 0.95 (Bentler, 1990). Finally, the RMSEA value of 0.033 is below the suggested maximum of 0.10 by Brunch (2008). With this evidence to support the model, it appears that the model is a good fit overall.

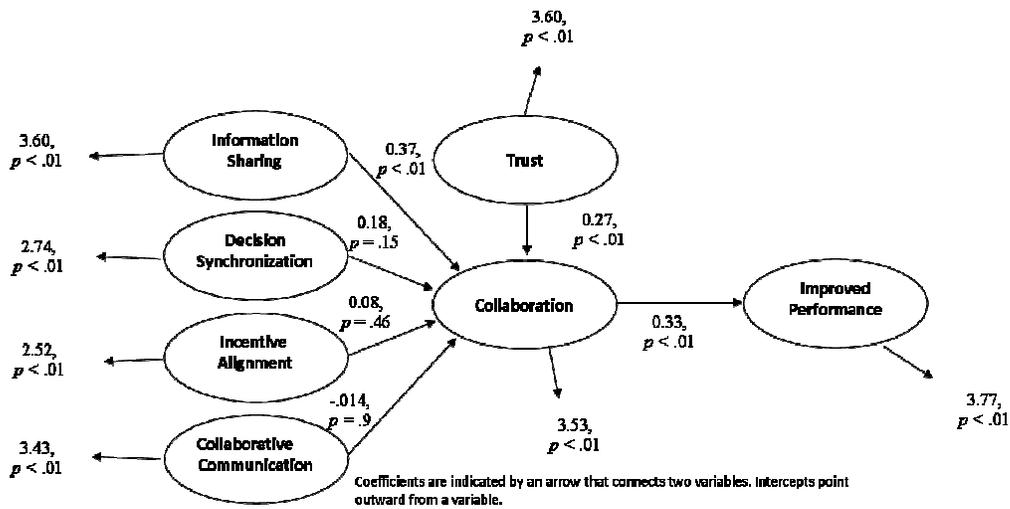

Figure 2. Measurement model

The measurement model results can be seen in Figure 2. It appears that information sharing has a positive and significant impact on collaboration, but the minimal relationship between decision synchronization and collaboration indicates that there is not a significant link between the two variables. In addition, trust appears to have a positive and significant relationship with collaboration.

Table 5. Hierarchical regression for collaboration continuum

| Variable | Model 1 | | | Model 2 | | |
|---|---|---|---|---|---|---|
| | Unstandardized B | SE B | Standardized Beta | Unstandardized B | SE B | Standardizea Beta |
| Firm Size | -0.081 | 0.089 | -0.159 | -0.067 | 0.054 | -0.132 |
| Annual Sales | -0.106 | 0.236 | -0.125 | -0.167 | 0.147 | -0.198 |
| Annual Purchasing | 0.13 | 0.215 | 0.145 | 0.135 | 0.132 | 0.151 |
| Contract | -0.010 | 0.339 | -0.004 | 0.200 | 0.211 | 0.071 |
| Information Sharing | | | | 0.389 | 0.099 | 0.363*** |
| Decision Synchronization | | | | 0.210 | 0.119 | 0.204* |
| Incentive Alignment | | | | 0.102 | 0.106 | 0.103 |
| Collaborative Communication | | | | 0.133 | 0.124 | 0.124 |
| Trust | | | | 0.189 | 0.087 | 0.200** |
| $R^2$ | 0.031 | | | 0.678 | | |
| F, Sig. | 0.579, 0.678 | | | 15.945, 0.000*** | | |

*** $p < .01$
** $p < .05$
* $p < .10$

In order to provide supporting evidence for the structural model findings, regression analyses were conducted using the factor scores associated with the relationships in Figure 2. Separate regressions were conducted to see the effects of each dimension and trust on collaboration, as well as collaboration on improved performance. Both models are statistically significant based upon their F statistics and consistent with the findings in the structural model. In addition, the VIF scores were examined for each variable and multicollinearity does not appear to be a problem since all scores were below 3.

Hierarchical regression analysis was also conducted on each model to account for the effects of covariates. These included firm size, annual sales volume, annual purchasing volume, and whether or not a formal agreement exists between respondent firms and any of their suppliers. The first regression considered the dimensions of collaboration and trust on collaboration. Results can be seen in Table 5. They indicate that the control variables had little impact on the regressions. Evidence includes a low model 1 $R^2$ value of 0.031, which increased to 0.678 for model 2. In addition, model 1 had an insignificant F value while model 2 was significant. Lastly, model 1 had no significant coefficients, but model 2 had three significant coefficients, including one at each of the $p < 0.01$, $p < 0.05$, and $p < 0.10$ levels.

The second hierarchical regression, which considered the impact of collaboration on performance, is presented in Table 6. It is again clear that the control variables had little to no impact on the results of the regressions. Evidence includes a low model 1 $R^2$ value of 0.019, which increased to 0.187 for model 2. In addition, model 1 had an insignificant F value while model 2 was significant. Lastly, model 1 had no significant coefficients, but model 2 had the Collaboration Continuum significant at the $p < 0.01$ level.

Table 6. Hierarchical regression for performance

| Variable | Model 1 | | | Model 2 | | |
| --- | --- | --- | --- | --- | --- | --- |
| | Unstandardized B | SE B | Standardized Beta | Unstandardized B | SE B | Standardized Beta |
| Firm Size | -0.010 | 0.078 | -0.022 | 0.021 | 0.072 | 0.047 |
| Annual Sales | -0.013 | 0.209 | -0.018 | 0.021 | 0.191 | 0.028 |
| Annual Purchasing | 0.124 | 0.190 | 0.157 | 0.085 | 0.174 | 0.108 |
| Contract | 0.166 | 0.300 | 0.027 | 0.070 | 0.275 | 0.028 |
| Collaboration Continuum | | | | 0.365 | 0.094 | 0.417** |
| $R^2$ | 0.019 | | | 0.187 | | |
| $F$, Sig. | 0.358, 0.837 | | | 3.367, 0.009** | | |
| | ** $p < .01$ | | | | | |

# 7. DISCUSSION

This research has investigated how much firms are collaborating with suppliers from the perspective of purchasing practitioners by utilizing an internet survey. The results provided evidence that collaboration is still not a fully utilized strategy for the respondent firms. More specifically, it does seem that firms have advanced beyond the traditional practice of combative relationships, but the conceptual ideal of collaboration does not yet seem to be commonplace.

Different dimensions of collaboration appear to have significantly different impacts on collaboration practice. It appears that firms engage in information sharing more frequently than other dimensions, and collaborative communication is engaged in more frequently than either decision synchronization or incentive alignment. An all-encompassing collaborative culture does not yet seem to be widespread, where firms can mutually benefit by reducing costs and inventory, and the final customer receives the best possible goods and services. This is consistent with the findings of Holweg et al (2005), who found that some collaboration initiatives are not being implemented as expected.

From a managerial perspective, it seems that relationships are critical components of any supply chain that do not yet receive adequate attention. Engaging in close relationships and being knowledgeable about the intricacies of other firms within its supply chain will give a firm the ability to be more adaptable and better suited to serve the end customer. While this should not be interpreted as a zero conflict situation, it does mean that situations in which obstacles arise must be handled appropriately.

Although it does seem to have a positive impact on collaboration, a key ingredient that seems to be missing from the survey results is interfirm trust. This is consistent with the findings of Fawcett et al (2012), who note that few firms have established a level of trust that will allow them to reap the benefits of collaboration. Since trust has been found to be a key driver to governance in collaborative relationships (Caniels et al, 2012), firms should strive to add this key ingredient to their relationships. Without this key aspect of collaboration, there will likely always be a significant barrier to any relationship.

This study makes it appear that collaboration still has significant room for improvement in buyer-supplier relationships, which may always be true to some degree. None of the results presented extraordinary evidence that collaboration is flourishing, which is consistent with findings of previous research, such as Daugherty et al (2006). However, as previously noted, not all partnerships require the relational intensity of a truly collaborative relationship. For example, a buyer will have a much different relationship with its supplier of custodial products than it does with a supplier of critical inputs. Thus, perhaps these results do not reflect poorly upon the effectiveness of a collaborative strategy, but simply reflect the varying nature of these relationships.

## 8. LIMITATIONS

While every effort was made to follow established guidelines in this research, it is in no way perfect and does suffer from weaknesses. This study only considers firms that are associated with the professional organization that sent the survey to its membership. Therefore, the study may suffer from a regional bias since it is restricted primarily to North America. Since many of the relationships examined in this research had no more than a national or even regional scope, the study may also not reflect the true nature of supply chain relationships in today's global economy since partnerships that span numerous national boundaries are no longer uncommon.

This research is also limited by the depth of information that can be captured by utilizing a survey as a primary methodological tool (Omar et al, 2012). In addition, the cross-sectional nature of the data collection method limits our perspective to a single point in time, so we are only able to view correlations between variables rather than causal effects. Our perspective is also limited by the fact that data was collected from only the purchasing perspective. It is quite possible that other viewpoints might provide varying results.

Another limitation to this study was the need for more respondents in order to increase the sample size. Due to a constrained budget for this study, e-mail was used as the sole means of distributing the survey to potential respondents. Kaplowitz et al (2004) found that conducting surveys on the web alone, or without supplemental mail reminders, led to the lowest response rate when compared to other methods, such as mail surveys. In addition, numerous sources indicate that e-mail open rates are quite low, ranging from the teens to roughly 30 percent. Thus, a majority of potential respondents may not even see the solicitation to participate in a survey, and only a small portion of those that see it follow through and participate. As a result, response rates for e-mail surveys may be lower than other data collection methods.

## 9. FUTURE RESEARCH

The research outlined above provides evidence about the benefits or detriments of collaboration. It is one ingredient to numerous other examples of research that seek to better understand the strategy. Firms and professionals can use this research to understand where they stand in terms of using collaboration as a business strategy and in terms of how they compare with other firms and supply chains. Academics can use the study to gain a better understanding of where collaboration truly stands in practice. The drive for innovation and race for publications has resulted in academia often describing a utopian collaboration situation that does not seem to be common, so this can aid researchers in knowing what is really happening rather than what will ideally happen in the perfect supply chain of the future.

An issue mentioned in more recent supply chain collaboration literature is power. Nyaga et al (2013) note that relationships do exist where there is a power balance between partners, but a more common occurrence is for relationships to have a power imbalance. This may have significant impacts on dimensions like decision synchronization and incentive alignment. It is generally the case that buyers have higher levels of power than sellers (Meehan and Wright, 2011), which has the potential to lead to the situation where one partner uses its power in an opportunistic way to gain more benefits (Sridharan and Simatupang, 2013). Since this clearly goes against the premise of collaboration, future research should identify the source of power imbalances and approaches to correct them.

In an effort to provide consistency in perspectives and make the results more meaningful, this research focused on the upstream portion of the supply chain by surveying purchasing representatives at respondent firms. Future research could expand this cross-sectional perspective into a more dyadic view. Investigating the opinions of sales managers, or those that are involved with the downstream portion of a firm's supply chain activities, could provide an alternative viewpoint that would supplement the results of this research. It may also be reasonable to investigate the opinions of both sides at the same time to see if their views conflict.

## 10. CONCLUSIONS

The results of this study provide evidence that the practice of supply chain management has not yet achieved its full potential in the procurement field. Firms seem to be maintaining a silo mentality that puts an emphasis on individual firm success rather than supply chain success. The results related to the supply chain continuum support this since firms seem to be talking about collaboration, but not actually implementing it. Boundary spanning activities, such as joint planning or synchronized decision making, are mentioned freely when purchasing professionals talk about their supplier relationships, but they do not seem to be widespread in practice. The data supports this since it has been found that information sharing seems to occur much more freely than collaborative communication, decision synchronization, or incentive alignment. Thus, firms are well on their way to practicing what the literature defines as collaboration, but this will only occur with improvements in these areas that are lacking.